# Scalable Zero-Knowledge Proofs for Verifying Cryptographic Hashing in Blockchain Applications

Oleksandr Kuznetsov [1,2,3], Anton Yezhov [4], Vladyslav Yusiuk [4], and Kateryna Kuznetsova [4,5]

[1] *Faculty of Engineering, eCampus University, Via Isimbardi 10, Novedrate (CO), 22060, Italy*
[2] *Department of Political Sciences, Communication and International Relations, University of Macerata, Via Crescimbeni, 30/32, 62100 Macerata, Italy*
[3] *Department of Information and Communication Systems Security, V. N. Karazin Kharkiv National University, 4 Svobody Sq., 61022 Kharkiv, Ukraine*
[4] *Zpoken OÜ, Harju maakond, Tallinn, Kesklinna linnaosa, Sakala tn 7-2, 10141, Estonia*
[5] *Department of Political Sciences, Communication and International Relations, University of Macerata, Via Crescimbeni, 30/32, 62100 Macerata, Italy*

**Abstract**

Zero-knowledge proofs (ZKPs) have emerged as a promising solution to address the scalability challenges in modern blockchain systems. This study proposes a methodology for generating and verifying ZKPs to ensure the computational integrity of cryptographic hashing, specifically focusing on the SHA-256 algorithm. By leveraging the Plonky2 framework, which implements the PLONK protocol with FRI commitment scheme, we demonstrate the efficiency and scalability of our approach for both random data and real data blocks from the NEAR blockchain. The experimental results show consistent performance across different data sizes and types, with the time required for proof generation and verification remaining within acceptable limits. The generated circuits and proofs maintain manageable sizes, even for real-world data blocks with a large number of transactions. The proposed methodology contributes to the development of secure and trustworthy blockchain systems, where the integrity of computations can be verified without revealing the underlying data. Further research is needed to assess the applicability of the approach to other cryptographic primitives and to evaluate its performance in more complex real-world scenarios.

## 1. Introduction

The advent of blockchain technology has revolutionized various industries, offering a decentralized and secure approach to data management and transactions [1]. However, as blockchain networks grow in size and complexity, scalability has emerged as a critical challenge. The increasing number of transactions and users on blockchain networks has led to slower transaction processing times and higher fees, hindering the widespread adoption of this technology [2]. Addressing the scalability issue is crucial for the success and practical applicability of modern blockchain projects.

Zero-knowledge proofs (ZKPs) have gained significant attention as a potential solution to the scalability problem in blockchain networks [3,4]. ZKPs allow one party (the prover) to prove to another party (the verifier) that a given statement is true without revealing any additional information beyond the validity of the statement itself [5,6]. By enabling the verification of transactions without disclosing sensitive data, ZKPs can significantly reduce the computational burden on blockchain nodes and improve the overall efficiency of the network [4,7]. The integration of ZKPs into blockchain systems has the potential to enhance privacy, security, and scalability, making them a promising tool for addressing the limitations of current blockchain implementations [3,8].

The primary objective of this research is to investigate the application of zero-knowledge proofs for ensuring the computational integrity of cryptographic hashing in blockchain systems. We aim to develop and evaluate a methodology for generating and verifying ZKPs using the Plonky2 framework, a state-of-the-art ZKP toolkit. The main tasks of this study include:
1. Generating ZKPs for cryptographic hashing of random data using Plonky2.
2. Testing the generated ZKPs to assess their correctness and efficiency.
3. Applying the developed methodology to real data blocks from the NEAR blockchain [9].
4. Analyzing the performance and scalability of the proposed approach for both random and real-world data.

By addressing these tasks, we seek to contribute to the development of efficient and scalable solutions for ensuring the integrity of computations in blockchain systems, ultimately supporting the broader adoption of this transformative technology.

## 2. Background

ZKPs are cryptographic protocols that allow a prover to convince a verifier of the validity of a statement without revealing any information beyond the truth of the statement itself [10]. The concept of ZKPs was first introduced by Goldwasser, Micali, and Rackoff in their seminal paper "The Knowledge Complexity of Interactive Proof Systems" [5]. A ZKP must satisfy three key properties:
1. Completeness: If the statement is true, an honest prover can convince an honest verifier of its validity.
2. Soundness: If the statement is false, no cheating prover can convince an honest verifier that it is true, except with a small probability.
3. Zero-knowledge: The verifier learns nothing beyond the truth of the statement.

Mathematically, a ZKP for a statement $x \in L$ can be represented as an interactive protocol between a prover $P$ and a verifier $V$, where $L$ is an NP language. The prover $P$ aims to convince the verifier $V$ that $x \in L$ without revealing any additional information. The protocol is described as follows [11]:

$$(P,V)(x) = \begin{cases} 1, \text{if } x \in L; \\ 0, \text{otherwise}, \end{cases}$$

where $(P,V)(x)$ denotes the output of the interaction between $P$ and $V$ on input $x$.

The construction of ZKPs relies on several cryptographic primitives and techniques, such as commitment schemes, challenge-response protocols, and hash functions [12]. A common approach to designing ZKPs is the Sigma protocol, which consists of three moves: commitment, challenge, and response [13,14].
1. Commitment: The prover sends a commitment to the verifier, which binds the prover to a specific value without revealing it.
2. Challenge: The verifier sends a random challenge to the prover.
3. Response: The prover computes a response based on the commitment, challenge, and the private information related to the statement being proved.

The verifier then checks the validity of the response and accepts or rejects the proof accordingly. The Fiat-Shamir heuristic [15] can be used to convert a Sigma protocol into a non-interactive ZKP by replacing the verifier's challenge with a hash of the prover's commitment and the statement being proved.

Cryptographic hash functions play a vital role in blockchain systems, ensuring the integrity and immutability of data. However, the computation of hash functions can be time-consuming, especially for large datasets. ZKPs can be employed to prove the correctness of hash computations without revealing the input data, thereby reducing the computational burden on blockchain nodes [16].

In this study, we focus on the application of ZKPs to the SHA-256 hash function, which is widely used in blockchain systems. The Plonky2 framework, developed by Polygon Zero, is utilized to generate and verify ZKPs for SHA-256 computations. Plonky2 is a modern ZKP toolkit that implements the PLONK protocol [17] in conjunction with FRI (Fast Reed-Solomon Interactive Oracle Proofs) [18] commitment scheme.

The PLONK protocol is a universally updatable structured reference string (SRS)

scheme that enables efficient proof generation and verification for arbitrary arithmetic circuits [17]. The FRI commitment scheme provides a fast and scalable method for committing to polynomials and verifying their evaluations [19].

By leveraging the capabilities of the Plonky2 framework, we aim to develop an efficient and scalable methodology for generating and verifying ZKPs for SHA-256 computations, ultimately contributing to the advancement of privacy-preserving and computationally efficient blockchain systems.

## 3. Research Methodology

In this section, we present a detailed description of the methodology employed for generating and testing ZKPs to ensure the computational integrity of cryptographic hashing. Our approach consists of three main stages: generating ZKPs for random data, testing the obtained proofs, and applying the developed methodology to real data blocks from the NEAR blockchain [9].

### 3.1. Generating ZKPs for Hashing Random Data

To generate ZKPs, we utilized the Plonky2 framework developed by Polygon Zero. Plonky2 implements the PLONK protocol in conjunction with FRI as a commitment scheme, providing a robust and efficient verification mechanism. The generation process involved the following steps:
1. Generating random data of various lengths (10, 100, 1000, 10000 bytes).
2. Computing the SHA-256 hash function for the generated data.
3. Creating a ZKP to validate the correctness of the hash computation using Plonky2.
4. Storing the generated proof and the corresponding circuit for subsequent verification.

The experiments for generating ZKPs were conducted on a server with an AMD Ryzen 9 7950X 16-Core Processor running at 4.7 MHz. For each length of random data (10, 100, 1000, 10000 bytes), we measured the following parameters:
- The complexity of native verification (computing the hash function and comparing the result with the hash code) in cycles per byte and seconds.
- The complexity of circuit generation in cycles per byte and seconds.
- The complexity of proof generation in cycles per byte and seconds.
- The complexity of proof verification in cycles per byte and seconds.
- The size of the generated circuit in gates.
- The size of the generated proof in bytes.

### 3.2. Testing the Generated ZKPs

To verify the correctness of the generated ZKPs, we employed the following methodology:
1. Loading the generated proof and the corresponding circuit for each set of random data.
2. Verifying the proof using Plonky2 while measuring the verification complexity in cycles per byte and seconds.
3. Comparing the hash code obtained from the verification with the original hash code computed for the random data.

### 3.3. Applying ZKPs to Real Data Blocks from the NEAR Blockchain

To assess the applicability of the developed methodology to real-world data, we utilized blocks from the NEAR blockchain of various heights and with different numbers of transactions:
- Block #121,114,606 at height 121,114,606, containing 52 transactions (5677 bytes) [20].
- Block #121,136,789 at height 121,136,789, containing 78 transactions (5092 bytes) [21].
- Block #121,117,653 at height 121,117,653, containing 102 transactions (4897 bytes) [22].
- Block #121,089,333 at height 121,089,333, containing 169 transactions (6262 bytes) [23].

The selected blocks reflect the diversity of real data in the NEAR blockchain and allow us to evaluate the performance of ZKPs generation and verification in various scenarios.

The process of generating and testing ZKPs for the selected NEAR blocks involved the following steps:
1. Obtaining the binary block data from the NEAR blockchain using the provided block hashes.

2. Generating a ZKPs for each block using Plonky2 while measuring the complexity of circuit and proof generation.
3. Verifying the generated proofs while measuring the verification complexity.
4. Comparing the obtained results with the results for random data to assess the applicability and scalability of the proposed approach.

By following this structured methodology, we aim to thoroughly evaluate the efficiency and practicality of generating and verifying ZKPs using Plonky2 for both random data and real data blocks from the NEAR blockchain [24]. The results of these experiments will be presented and discussed in the following section.

## 4. Results and Analysis

In this section, we present and analyze the results obtained from generating and testing ZKPs for both random data and real data blocks from the NEAR blockchain. The experiments were conducted using the methodology described in the previous section, and the results provide valuable insights into the efficiency and scalability of the proposed approach.

### 4.1. Results for Random Data

Table 1 summarizes the results of generating and testing ZKPs for random data of various lengths using the Plonky2 framework. The table includes the complexity of native verification, circuit generation, proof generation, and proof verification, as well as the sizes of the generated circuits and proofs.

The results in Table 1 demonstrate the following key observations:
1. The complexity of native verification, circuit generation, proof generation, and proof verification increases with the length of the random data. However, the increase in complexity is not linear, indicating the scalability of the proposed approach.
2. The time required for native verification remains negligible (in the order of microseconds) even for larger data lengths, highlighting the efficiency of the native verification process.
3. The time required for circuit generation and proof generation increases with the data length, but remains within acceptable limits (less than 13 seconds for 10000 bytes of data).
4. The time required for proof verification is significantly lower than that of proof generation, emphasizing the efficiency of the verification process, which is crucial for the practical application of ZKPs.
5. The sizes of the generated circuits and proofs increase with the data length, but remain manageable (less than 250 KB for 10000 bytes of data), ensuring the feasibility of storing and transmitting the generated proofs.

These results confirm the efficiency and scalability of the proposed approach for generating and verifying ZKPs using the Plonky2 framework for random data of various lengths.

To illustrate the relationship between the complexity of proof verification and the length of random data, we present Figure 1, which shows the proof verification time as a function of the input data size.

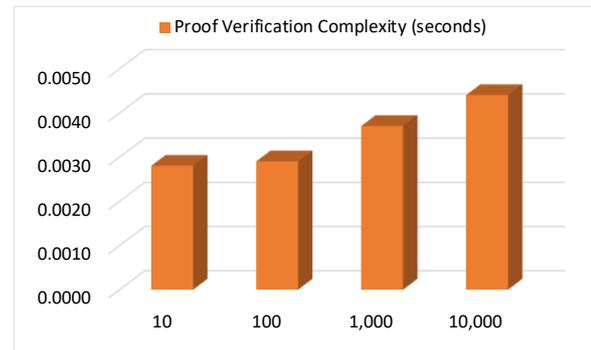

**Figure 1:** Proof verification time as a function of random data length

As evident from Figure 1, the proof verification time remains consistently low, even for larger data sizes, with a verification time of approximately 0.0044 seconds for random data of length 10,000 bytes. This observation highlights the efficiency of the verification process and its potential for scalability in real-world applications.

The linear relationship between the verification time and the data length can be attributed to the design of the Plonky2 framework and the underlying cryptographic primitives, such as the PLONK protocol and the FRI commitment scheme. The efficient arithmetic circuit representation and the optimized proof construction in Plonky2 contribute to the fast verification times, even for larger datasets.

**Table 1**
Results of generating and testing ZKPs for random data of various lengths

| Random Data Length (bytes) | 10 | 100 | 1,000 | 10,000 |
|---|---|---|---|---|
| Native Verification Complexity (cycles/byte) | 196 | 250 | 1,752 | 17,022 |
| Native Verification Complexity (seconds) | 0.00000004 | 0.00000005 | 0.0000003 | 0.000003 |
| Circuit Generation Complexity (cycles/byte) | 197,896,186 | 432,109,936 | 5,653,509,584 | 58,641,652,143 |
| Circuit Generation Complexity (seconds) | 0.04 | 0.09 | 1.23 | 12.70 |
| Proof Generation Complexity (cycles/byte) | 255,670,545 | 465,505,001 | 3,730,111,317 | 38,720,856,965 |
| Proof Generation Complexity (seconds) | 0.05 | 0.10 | 0.82 | 8.58 |
| Proof Verification Complexity (cycles/byte) | 11,826,688 | 12,459,491 | 15,544,539 | 19,610,437 |
| Proof Verification Complexity (seconds) | 0.0028 | 0.0029 | 0.0037 | 0.0044 |
| Circuit Size (gates) | 1,419 | 2,842 | 22,739 | 223,148 |
| Proof Size (bytes) | 121,752 | 127,256 | 152,756 | 180,112 |

**Table 2**
Results of generating and testing ZKPs for real data blocks from the NEAR blockchain

| Block Height | 121,114,606 | 121,136,789 | 121,117,653 | 121,089,333 |
|---|---|---|---|---|
| Number of transactions | 52 | 78 | 102 | 169 |
| Block bytes | 5,677 | 5,092 | 4,897 | 6,262 |
| Native Verification Complexity (cycles/byte) | 9,368 | 8,424 | 8,366 | 10,318 |
| Native Verification Complexity (seconds) | 0.000001 | 0.000001 | 0.000001 | 0.000002 |
| Circuit Generation Complexity (cycles/byte) | 27,010,322,753 | 26,380,158,107 | 26,791,174,445 | 56,830,642,601 |
| Circuit Generation Complexity (seconds) | 5.87 | 5.71 | 5.80 | 12.18 |
| Proof Generation Complexity (cycles/byte) | 18,633,537,207 | 19,172,519,712 | 18,015,459,404 | 38,191,422,267 |
| Proof Generation Complexity (seconds) | 4.18 | 4.10 | 4.03 | 8.32 |
| Proof Verification Complexity (cycles/byte) | 17,173,339 | 17,197,238 | 17,034,388 | 19,024,157 |
| Proof Verification Complexity (seconds) | 0.004 | 0.004 | 0.004 | 0.004 |
| Circuit Size (gates) | 126,498 | 113,704 | 109,442 | 139,289 |
| Proof Size (bytes) | 165,684 | 165,684 | 165,684 | 180,112 |

## 4.2. Results for Real Data Blocks from the NEAR Blockchain

To assess the applicability of the developed methodology to real-world scenarios, we generated and tested ZKPs for data blocks from the NEAR blockchain. Table 2 presents the results obtained for the selected blocks, including the block height, the number of transactions, the block size, and the complexity and time required for native verification, circuit generation, proof generation, and proof verification.

The results in Table 2 lead to the following observations:
1. The complexity of native verification for real data blocks is comparable to that of random data of similar sizes, confirming the consistency of the native verification process.
2. The time required for circuit generation and proof generation for real data blocks is also comparable to that of random data, demonstrating the applicability of the proposed approach to real-world scenarios.
3. The time required for proof verification remains consistently low (around 0.004 seconds) for all the tested real data blocks, regardless of the number of transactions or block size, highlighting the efficiency of the verification process.
4. The sizes of the generated circuits and proofs for real data blocks are similar to those of random data, indicating the feasibility of storing and transmitting the proofs in real-world applications.

These results validate the applicability and scalability of the proposed methodology for generating and verifying ZKPs using Plonky2 for real data blocks from the NEAR blockchain.

The consistency in performance between random and real data suggests that the approach can be effectively utilized in practical scenarios, such as ensuring the computational integrity of cryptographic hashing in blockchain applications.

## 5. Discussion

The experimental results presented in this section demonstrate the efficiency and scalability of the proposed approach for generating and verifying ZKPs using the Plonky2 framework. The methodology exhibits consistent performance for both random data and real data blocks from the NEAR blockchain, highlighting its potential for practical applications.

The complexity of native verification, circuit generation, proof generation, and proof verification scales well with increasing data lengths, ensuring the feasibility of applying the approach to larger datasets. The time required for proof verification remains consistently low, even for real data blocks with a large number of transactions, emphasizing the efficiency of the verification process, which is crucial for the practical adoption of ZKPs.

Moreover, the sizes of the generated circuits and proofs remain manageable, even for larger data lengths and real data blocks, indicating the feasibility of storing and transmitting the proofs in real-world scenarios. This is particularly important for blockchain applications, where the storage and transmission of proofs should not introduce significant overhead.

The consistency in performance between random and real data suggests that the proposed methodology can be effectively applied to ensure the computational integrity of cryptographic hashing in various applications, including blockchain systems. The ability to generate and verify ZKPs efficiently and scalably can contribute to the development of more secure and trustworthy systems, where the integrity of computations can be verified without revealing the underlying data.

However, it is important to note that the current study focuses on the specific case of cryptographic hashing using the SHA-256 algorithm. Further research is needed to assess the applicability of the proposed approach to other cryptographic primitives and to evaluate its performance in more complex real-world scenarios.

In conclusion, the experimental results presented in this section provide strong evidence for the efficiency and scalability of the proposed methodology for generating and verifying ZKPs using the Plonky2 framework. The approach demonstrates consistent performance for both random and real data, highlighting its potential for practical applications in ensuring the computational integrity of cryptographic hashing, particularly in the context of blockchain systems.

## 6. Conclusions

In this study, we proposed and evaluated a methodology for generating and verifying ZKPs to ensure the computational integrity of cryptographic hashing in blockchain systems. By leveraging the Plonky2 framework, we demonstrated the efficiency and scalability of our approach for both random data and real data blocks from the NEAR blockchain.

The experimental results showed that the proposed methodology achieves consistent performance across different data sizes and types, with the time required for proof generation and verification remaining within acceptable limits. The complexity of native verification, circuit generation, proof generation, and proof verification scales well with increasing data lengths, indicating the feasibility of applying the approach to larger datasets.

Moreover, the sizes of the generated circuits and proofs remain manageable, even for real-world data blocks with a large number of transactions. This is particularly important for blockchain applications, where the storage and transmission of proofs should not introduce significant overhead.

The consistency in performance between random and real data suggests that the proposed methodology can be effectively applied to ensure the computational integrity of cryptographic hashing in various blockchain systems. The ability to generate and verify ZKPs efficiently and scalably contributes to the development of more secure and trustworthy systems, where the integrity of computations can be verified without revealing the underlying data.

Further research is needed to assess the applicability of the proposed approach to other cryptographic primitives and to evaluate its performance in more complex real-world scenarios. Nonetheless, the results presented in this study provide a solid foundation for the development of efficient and scalable solutions for ensuring the integrity of computations in blockchain systems, ultimately supporting the broader adoption of this transformative technology.